\documentclass[%
 prl,
 preprint,
 superscriptaddress,
nobibnotes,
 amsmath,amssymb,
 aps,
floatfix,
noeprint
]{revtex4-2}

\usepackage{graphicx}
\usepackage{siunitx}
\usepackage{ulem} 
\usepackage[utf8]{inputenc}
\usepackage{xspace}
\usepackage{afterpage}

\renewcommand{\rm}[1]{{}}

\graphicspath{{./Figures}}

\begin{document}

\title{Dimensional crossover in a quantum gas of light }

\author{Kirankumar Karkihalli Umesh\footnote{ORCID: 0000-0002-3644-1233}}
\affiliation{Institut für Angewandte Physik, Universität Bonn, Wegelerstrasse 8, 53115 Bonn, Germany}

\author{Julian Schulz\footnote{ORCID: 0000-0003-4630-4117}}
\affiliation{Physics Department and Research Center OPTIMAS, RPTU Kaiserslautern Landau, 67663 Kaiserslautern, Germany}

\author{Julian Schmitt\footnote{ORCID: 0000-0002-0002-3777}}
\affiliation{Institut für Angewandte Physik, Universität Bonn, Wegelerstrasse 8, 53115 Bonn, Germany}

\author{Martin Weitz\footnote{ORCID: 0000-0002-4236-318X}}
\affiliation{Institut für Angewandte Physik, Universität Bonn, Wegelerstrasse 8, 53115 Bonn, Germany}

\author{Georg von Freymann\footnote{ORCID: 0000-0003-2389-5532}}
\affiliation{Physics Department and Research Center OPTIMAS, RPTU Kaiserslautern Landau, 67663 Kaiserslautern, Germany}
\affiliation{Fraunhofer Institute for Industrial Mathematics ITWM, 67663 Kaiserslautern, Germany}

\author{Frank Vewinger\footnote{ORCID: 0000-0001-7818-2981}\footnote{Corresponding Author, vewinger@uni-bonn.de}}

\affiliation{Institut für Angewandte Physik, Universität Bonn, Wegelerstrasse 8, 53115 Bonn, Germany}

\date{\today}

\begin{abstract}


{The dimensionality of a system profoundly influences its physical behaviour, 
leading to the emergence of different states of matter in many-body quantum 
systems. In lower dimensions, fluctuations increase and lead to the suppression of 
long-range order. For example, in bosonic gases, Bose-Einstein condensation in 
one dimension requires stronger confinement than in two dimensions. Here, we 
observe the dimensional crossover from one to two dimensions in a harmonically 
trapped photon gas and study its properties. The photons are trapped in a dye 
microcavity where polymer nanostructures provide the trapping potential for the 
photon gas. By varying the aspect ratio of the harmonic trap, we tune from an 
isotropic two-dimensional confinement to an anisotropic, highly elongated one
dimensional trapping potential. Along this transition we determine the caloric 
properties of the photon gas and find a softening of the second-order Bose
Einstein condensation phase transition observed in two dimensions to a crossover 
behaviour in one dimension. }

\end{abstract}

\maketitle

\section{Main Text}

In the world of manybody physics, it is common knowledge that the number of accessible dimensions profoundly influences the physical behaviour of a system, leading to the emergence of different states of matter at low dimensions ({\itshape{i.e.}}, less than three dimensions). For bosonic gases, as an example, Bose-Einstein condensation (BEC) is possible in lower dimensions only for a sufficiently strong confinement of a power-law trapping potential~\cite{Bagnato.1991}. While in two dimensions (2D) a harmonically trapped gas can undergo a phase transition to a Bose-Einstein condensate at finite temperature, this transition in one dimension (1D) is only observed with tighter confining power-law traps. When confining a 1D system within a harmonic trap there is no sharply defined phase transition in the thermodynamic limit to the condensate phase \cite{Ketterle.1996}, but rather a smooth crossover to a quasi-condensate. In this regime, large thermal and quantum fluctuations in one dimension inhibit the emergence of true long-range order~\cite{Dettmer.2001,Kruger.2010}. For finite-size systems, the change from a phase transition in 2D to a crossover in 1D is less pronounced.  Moreover, condensation can occur in 1D depending on the strength of interactions and different regimes for quantum-degenerate Bose gases are expected~\cite{Petrov.2000}. It is therefore of great interest to study the crossover from three-dimensional (3D) or two-dimensional (2D) systems to the 1D case. For ultracold atomic gases, both thermodynamic and coherence properties associated with the emergence of quasi-long range order along the \textit{dimensional crossover} from three to one dimension have been studied experimentally~\cite{Vogler.2014,Shah.2023}. In the case of a strongly interacting Bose gas, where one expects a crossover from Berezinski-Kosterlitz-Thouless type correlations in 2D to Tomonaga-Luttinger liquid correlations \cite{Yao_2023},  the interplay between interactions and dimensionality has been studied \cite{guo_2023}, and it has been observed that reducing the dimensionality can strongly influence the temperature of the system \cite{guo_2023cooling}. For dipolar gases, the transition to the supersolid phase has been studied along the dimensional crossover from one to two dimensions,  showing  a transition from a continuous to a discontinuous behaviour {in the order parameter} \cite{Biagioni.2022}. 

Optical quantum gases have in recent years emerged as an alternative platform for quantum gas experiments~\cite{Carusotto.2013}, which are well-suited for the study of the dimensional crossover from 2D to 1D due to the weak or even negligible interaction. In exciton-polariton condensates, correlations in 1D, 2D and 3D have been studied theoretically for an interacting gas~\cite{Chiocchetta.2013}. Here, one crosses from long-range order in 3D via a power law in 2D to exponentially decaying {first-order phase} correlations in the 1D case. Experimentally, e.g.~the formation dynamics has been studied for 1D systems \cite{Wertz.2010,Wouters.2010}, and in 1D coupled lattices, a Kardar--Parisi--Zhang {(KPZ)} scaling for the phase has been observed \cite{Fontaine.2022}. {In a semiconductor microcavity the transition from 2D to 1D has been studied by varying the geometry of the drive, where the dissipative phase transition observed in 2D vanishes for a 1D drive \cite{Li2022}. }

In weakly- or non-interacting photon {gases}, thermalization and condensation in {1D} has been observed in doped optical fibers, where the photons {thermally populate} the longitudinal degree of freedom {in a long single mode fibre~\cite{Weill.2021}}. To achieve condensation, the initially linear dispersion relation is altered to a sub-linear dispersion using chirped gratings, together with a spectral filter to provide a nontrivial ground state {at finite energy}. In contrast to this, one can confine photons in a microcavity, effectively freezing out the longitudinal degree of freedom~\cite{Klaers.2010}. The transverse degrees of motion can be restricted by {in-plane} trapping potentials induced by transverse variations in the optical path lengths, either by structuring the cavity mirror surface~\cite{Kurtscheid.2020,Walker:21} or the local refractive index~\cite{Dung.2017,Vretenar.2021}. By this, {variable} potential {geometries} can be {realized}, which have{, \textit{e.g.},} allowed to study the thermodynamics of {2D Bose gases which contain a} few photons {or are trapped in} box or double well potentials~\cite{Walker.2018,Busley.2022,Kurtscheid.2019}. For this system, a continuous change in the thermodynamic properties has been predicted for the harmonically trapped gas when crossing from {a 2D to 1D configuration}~\cite{Stein.2022,Stein.2022a}, where in contrast to the 2D case no Bose-Einstein condensation is expected \cite{Bagnato.1991}.

Here, we use a novel technique to confine photons, namely by printing polymer structures on top of one of the cavity mirrors, which allows us to prepare structures with sizes below the wavelength of the photons. Using this method, we study the transition from two to one dimension for a harmonically trapped gas of photons by varying the aspect ratio of the trapping potential. In our system, thermalization occurs via radiative contact of the photons to a bath of dye molecules, and correspondingly the thermalization {mechanism is decoupled from the} dimensionality of the trapping potential, in contrast to atomic Bose gases~\cite{Kinoshita.2006,Mazets.2008}. For all {investigated} aspect ratios a macroscopic occupation of the ground state is observed {as the} photon number {is increased}, which in two dimensions is accompanied by a sharp transition in the chemical potential, while in one dimension we observe a smooth crossover. For the intermediate cases, we observe a {gradual} softening of the phase transition, which can be associated to an effective (non-integer) dimension of the system. 

We prepare our photon gas in an optical microcavity consisting of two highly reflective {plane} mirrors (reflectivity above 99.995\% at \SI{580}{\nano\meter}) spaced by approximately \SI{2}{\micro\meter} filled with a dye solution, see Fig.~1a. The confinement in longitudinal direction effectively freezes out this degree of freedom, yielding a minimum energy of $\hbar\omega_c\approx \SI{2.1}{\electronvolt}$ for the photons in the cavity, corresponding to a wavelength of \SI{580}{\nano\meter}. To confine the photons in the transversal direction we printed a polymer micro\rm{-}structure of height $d{(x,y)}$ on one of the flat mirrors. The refractive index of the polymer $n_{s}$ {exceeds} the refractive index $n$ of the dye solution, such that the optical path length locally increases, leading to an attractive potential energy landscape for the photons {in the regions where the polymer is deposited} with a potential  $V \propto d{(x,y)} (n - n_{s})/n$ \cite{Vretenar.2021}. The geometry of the potential is determined by the geometry of the printed structure, $d{(x,y)}$, where the {surface} curvature of the printed structure translates to the curvature of the trapping potential and the {absolute} height to {its} trap depth. For the used {dye} solvent ethylene glycole ($n \approx 1.44$) and polymer ($n_{s}\approx 1.55$)~\cite{Gissibl:17}, we find a {trap} depth $V \approx 1.27 k_\mathrm{B} T$ for the used {maximum} structure height of approximately \SI{325}{\nano\meter}. The polymer nanostructuring allows us to fabricate parabolic structures with {sufficiently strong} curvatures in the tightly-confining $y$-direction, along which the corresponding harmonic oscillator potential contains only a single bound state, rendering the photon gas system effectively one-dimensional (see Fig.~1b).

The polymer structures were fabricated out of the negative-tone photoresist IP-Dip by a direct laser writing (DLW) system (Nanoscribe Photonic Professional GT) \cite{Hohmann.2015}. The substrate was prepared in the immersion configuration, which means that during the writing process, the laser was focused by the objective first through an immersion medium (in our case again IP-Dip) and the substrate of the cavity mirror and at last through the thin dielectric Bragg layers into the photoresist (see Fig.~1c). The writing trajectory followed parallel lines along the longer axis of the potentials ($x$-direction) with a line distance of \SI{100}{\nano\meter} (along the $y$-direction). To manufacture the desired potential {we print a polymer ridge with height profile }$d(x,y)=h_0-c_x x^2-c_y y^2$ , where $h_0$ {denotes} the maximum height of the structure and $c_x$ and $c_y$ the curvatures along $x$- and $y$-direction, respectively, at each point the polymer was exposed up to a height given by $d(x,y)$. For the different potentials, we keep the geometry approximately fixed along the $x$-direction and vary the curvature along the $y$-direction. In the following, we {label} the {different} potentials by their aspect ratio $\Lambda=\omega_y/\omega_x=\sqrt{c_y/c_x}$, with the trap frequencies $\omega_x$ and $\omega_y$. The ratio $\Lambda$ {quantifies the effective} dimension of the photon gas \cite{Stein.2022}, where $\Lambda =1$ corresponds to an isotropic 2D harmonic oscillator with equal trapping frequency along $x$- and $y$- directions, {see Extended Data Fig. 1}, and quasi-1D is reached when the first excited mode of the strongly confined dimension is not trapped in the potential anymore (which in our case also implies $k_B T < \hbar \omega_y$), which in our case is achieved at $\Lambda \approx 22$. 

To thermalize the photons they are coupled to a thermal bath at ambient temperature $T=\SI{300}{\kelvin}$, realised by a dye solution filled between the cavity mirrors (see Fig.~1a), similar to previous work \cite{Klaers.2010,Klaers.2011}. By repeated absorption-emission cycles photons thermalize to the temperature of the dye solution provided thermalization is sufficiently faster than the photon losses, as is the case in our system~\cite{Kirton.2015,Klaers.2011,Schmitt.2015}. Correspondingly, the photons populate the energy levels of the transversal degrees of freedom, {\textit{i.e.},} the harmonic oscillator levels in $x$- and $y$-direction, leading to a spectrum with equidistant {frequency} spacing above the lowest energy mode ({called the 'cutoff'} energy or frequency)~\cite{Klaers.2010}. {As the thermalization is achieved by coupling to a bath and not by direct photon-photon collisions, we even for a few photons expect a thermal distribution, and correspondingly for strongly confining 2D potentials Bose-Einstein condensates with less than 10 photons have been reported~\cite{Walker.2018,Kurtscheid.2019}.} To prepare the initial photon population and to compensate for losses out of the system we exploit the low reflectivity of our mirrors at \SI{532}{\nano \meter} to inject dye molecular excitation using a laser at \SI{532}{\nano \meter}, {which fixes the chemical potential of the photons and thus the average total photon number $N$}. The pump light is time modulated with \SI{500}{\nano\second} pulse width at \SI{50}{\hertz} repetition rate to prevent bleaching of the dye molecules \cite{Klaers.2010}. A spatial light modulator \rm{(SLM)}is used to shape the pump laser profile to match the structure size on the mirror, thus reducing unwanted fluorescence from unconfined modes from outside the {polymer-based} structure.

To analyse the photon gas we collect the light emitted through one of the cavity mirrors, and split the transmitted radiation into two paths after lifting the polarization degeneracy using a polarizer {oriented along the polarization direction of the pump radiation, which coincides with the long axis of the 1D potential}. About 70\% of the light is collected by a spectrometer, and 30\% are used for spatial imaging of the photon gas. Typical observed density distribution of the photon gas trapped in the 1D to 2D potential are shown in Fig~2, {and momentum space images are depicted in Extended Data Fig. 2.}
In the spatial distributions, one can see the density {tightly} squeezed along the $y$ direction for 1D ($\Lambda =22$), elliptic for the 1D-2D ($\Lambda=5$) potential, and radially symmetric for the isotropic 2D potential ($\Lambda=1$). The profiles well follow expectations given by a Bose-Einstein distributed population within the bound modes of the harmonic oscillator potential. {All images show data taken in the quantum degenerate regime, with a macroscopic population of the ground mode. To visualize this, the theory expectations (see Methods) for ground mode (red) and thermal modes (black dashed) are shown along with the cuts.} Especially for the 1D potential, the finite size of the potential becomes visible in the diffraction pattern on the sides of the emission. {Additionally, one observes a broader smeared out background below the sharp peak originating from trapped photons, which is attributed to residual fluorescence from free space modes above the harmonic oscillator potential, and also to pump light scattered at the edges of the potential, which leads to an increase in emission at the edge of the potential. Those features are also present in the other cases but can be neglected due to the larger photon numbers in the potential.} As these modes can be separated spectrally, in the following we focus on the spectroscopy of the cavity emission to study the caloric properties of the gas. For this, the emission is dispersed energetically along the $y$-direction using an optical grating and imaged onto an sCMOS camera (called \textit{raw spectra} in the following), which allows us to measure both the population in and the spatial profile of individual modes simultaneously. {Our slitless spectrometer, which is described in more detail in the online methods, has a spectral resolution of $\approx$\SI{0.08}{\tera\hertz} for the lowest modes.} For higher modes with quantum numbers $n_y\gg 1$ the modes start to spatially overlap, correspondingly lowering the resolution for highly excited modes. We extract a spectrum by integrating the obtained raw spectra along the non-dispersed direction, {averaging over 30 realizations for a specific total photon number in 2D, over 90 in the 2D-1D case and 120 for 1D data.}

Exemplary measured spectra are shown in Fig.~3{a} for three different aspect ratios, together with the {corresponding} raw spectra in Fig.~3b obtained by dispersing the cavity emission using a grating, which retains the mode profile along the x-direction. For {the} isotropic 2D harmonic oscillator potential {with $\Lambda=1$} {(top panel)}, one observes equidistantly spaced modes with a {frequency} spacing of \SI{0.223}{\tera \hertz}. In the raw spectra, {individual modes $(n_x,n_y)$ with $n_x+n_y=const$ spatially overlap, as harmonic oscillator modes with equal $n_x+n_y$ have the same energy. Correspondingly, the TEM00 mode can be distinguished from the group of modes TEM01 and TEM10, but the latter two overlap. This can be seen more clearly in a linear color scale (see Extended Data Fig. ), however there the thermal part is not visible any more.} For $\Lambda =5$, corresponding to an anisotropic 2D harmonic oscillator potential with $\omega_y = 5\omega_x$ (middle panel), one observes equidistantly spaced modes except for the few lowest modes due to slight distortions in the polymer structure. The mode degeneracy increases every 5th mode, as can be seen by the step-like increase in intensity at those modes. In the raw spectrum, this is reflected by the emergence of multiple parabolas corresponding to the different quantum numbers $n_y$ along the strongly confined $y$-direction. One correspondingly finds the emergence of the second dimension ($n_y=1$) around the 5th mode ($n_x=5$) along the relaxed $x$-direction.
The potential with $\Lambda =22$ displays 1D harmonic oscillator modes {(bottom panel)} in the raw spectrum, with an energy spacing of \SI{0.37}{\tera \hertz}, {and} the integrated spectrum {correspondingly} shows discrete peaks with a degeneracy of one, see also Extended Data Fig.~3. This is characteristic for the absence of the second dimension in the recorded energy interval and which is confined by the trap for the strongly asymmetric harmonic oscillator potential, {demonstrating that a photon gas trapped in} this potential can be considered as being effectively 1D. The photon distribution in all three cases well follows the Bose-Einstein distribution, the black markers {in Fig.~3a} indicate the expected photon distribution when neglecting the width of the individual modes. The theory estimations were calculated using a Bose-Einstein distribution $g(E)  \left(\text{e}^{(E-\mu ) / k_{B}T} -1\right)^{-1}$, with the degeneracy $g(E)$. We use the energies of a quantum harmonic oscillator spectrum, $E = \hbar \left[ \omega_{x} n_x + \omega_{y} n_y+ \frac{1}{2}(\omega_{x}+\omega_{y}) \right] $. 
The trap frequencies $\omega_{x}$ and $\omega_{y}$ are extracted from the mean mode spacing of the measured spectra, and we truncate the theory spectra by excluding the energy levels which exceed the trap depth inferred from the measured spectrum, and the photon number at each mode is determined by calibrating the signal of the sCMOS camera. The experimental data well follow the theory expectations, apart from a slightly lower mode population for the 2D case. When including the mode profile of individual modes the measured spectrum also in this case well follows expectation for a thermal distribution at 300\,K, see Extended Data Figs.~4 and 5 the Methods for details.

Using the measured spectra, we extract the photon number in the ground mode and the excited modes, respectively, as a function of the total photon number as shown in Fig.~4. Here, one clearly observes a smooth crossover in the ground state population for the 1D case, as expected as in 1D no phase transition to a BEC occurs. The transition becomes sharper for $\Lambda=5$, and {shows the steep increase which is associated with} the  phase transition for the isotropic 2D potential with $\Lambda =1$. In the latter case, we also observe the saturation of excited modes, as expected for a phase transition from a thermal gas to a Bose-Einstein condensate. For the 1D case, the number of states bound in the potential is smaller than in the 2D case, and correspondingly the softening of the phase transition when crossing from 2D to 1D might also indicate a finite size effect. To investigate this, Fig.~4d shows the theory expectations for a 2D harmonic potential with the same number of energy levels as in the 1D case. While one observes a softening due to finite size effects, the effect is smaller than for the transition to 1D, and correspondingly the experimental data gives evidence for the dimensional crossover from 1D to 2D. This is also visible  in Fig.~3a, where no macroscopic population in the ground mode is visible.

The influence of the dimension on the phase transition can be explored by studying the order parameter when tuning the system parameters. As for our case the temperature is fixed at room temperature, the tuning parameter is the total photon number, and we use the absolute value of the chemical potential $|\mu|$ as an order parameter. To find the chemical potential $\mu$, we first extract the internal energy from spectra as in Fig.~3 for different total photon numbers. For each spectrum, we set the ground mode energy to be the zero point energy of the harmonic oscillator, $E_{0} =\hbar(\omega_{x} + \omega_{y})/2 $, count the number of photons in each energy level, multiply by the corresponding mode energy and sum over the whole spectrum, yielding the internal energy as shown in Fig.~5b. For all shown aspect ratios, the internal energy per photon, $U/N$, decreases for increasing photon numbers indicating an increasing population in the low-energy states. As we don't expect a sharp phase transition in 1D (and thus no well-defined critical photon number), for better comparison, each data set is scaled to the photon number $\tilde{N}$ where the chemical potential $\mu(N)$ equals half the chemical potential at low photon number, $\mu(\tilde{N}) = \mu(N\rightarrow 0)/2$, i,e, halfway between the chemical potential  for the  classical gas and the quantum degenerate case. We extract this number from the theoretical curves based on Bose-Einstein distributed occupations within modes {(see the online Methods {and Extended Data Fig. 6} for details)},  which yields $\tilde{N}=$ 628, 64 and 23 photons for the 2D, the 2D-1D and the 1D harmonic oscillator potentials respectively. In the isotropic 2D potential ($\Lambda = 1$), the curve changes slope sharply around $N/\tilde{N}\approx 1$, indicating the presence of phase transition, while in the 1D potential ($\Lambda=22$), the slope changes monotonically, and shows the absence of a {thermodynamic phase transition}. As for the anisotropic 2D potential ($\Lambda=5$), the $U/N$ slope indeed has, although less strong, a sharp change as in 2D isotropic potential Fig.~5.

This is visible more strongly in the chemical potential, by numerically taking the partial derivative of the internal energy $U$ with respect to the photon number $N$. The numerical derivative was done by first binning the photon number data for $U$ (bins of photon number in a geometric series spacing with a common ratio of 1.2, 1.3 and 1.2 for the 2D, the 2D-1D and the 1D harmonic oscillator potentials respectively) to suppress {numerical noise}{, see Extnded Data Figs.~7--9 for the influence of the binning.} The absolute value of $\mu$ as expected decreases for all three aspect ratios with increasing photon number. The sharp drop followed by the saturation at the ground state energy for the isotropic 2D case $\Lambda =1$ indicates the expected phase transition in two dimensions, while for increasing $\Lambda$ we observe a gradual softening of the transition to a continuous crossover for the 1D potential. Thus, the change in the dimension of the potential from 2D to 1D leads to a crossover between the different regimes of the Bose gas instead of a sharp phase transition to a condensate phase.

To conclude, we have experimentally studied the dimensional crossover from a {2D} isotropic harmonically trapped photon gas to a photon gas confined to {1D} around the transition from a thermal to the quantum degenerate case. This crossover is accompanied by a softening of the phase transition, which crosses from a true {second-order} phase transition to a BEC in {2D}  to a continuous behaviour in {1D}, indicated by the behaviour of the chemical potential {and the internal energy of the photon gas} for different photon numbers.

For the future, it will be interesting to study different trapping potentials for the photons, and investigate the spatial correlations~\cite{Damm.2017}. While in 1D for the harmonically trapped system no true long-range order is possible, in our finite-size system the correlations can extend over the whole system. In the cavity platform the losses can be tuned from a nearly lossless system to the case of a driven-dissipative condensate by modifying the low-frequency cutoff~\cite{Oeztuerk.2021}, which is expected to alter the correlations in the system, and additionally influence the polarization properties of the condensate \cite{Moodie.2017}. Also, the structuring method presented here, based on polymer structures within the cavity,  allows great flexibility in the design of potentials for photons, ranging from continuous potentials like the ones presented here to tunnel-coupled lattice structures with large tunneling rates. For example, potentials with a logarithmic level spacing have been proposed for factorization of large numbers \cite{Gleisberg_2013}, and in tunnel-coupled potentials, the influence of loss and drive e.g. leads to stable vortices~\cite{Gladilin.2020,Wouters.2022}, clustering~\cite{Panico.2023} or the emergence of a {KPZ}-like scaling in the correlations~\cite{Fontaine.2022}, and for 1D chains the emergence of surface states is possible in the presence of a retarded thermooptic interaction~\cite{Strinati.2022}.

\textbf{Acknowledgements} \par 
We acknowledge valuable discussions with E. Stein and A. Pelster. This work has been supported  by the Deutsche Forschungsgemeinschaft through CRC/Transregio 185 OSCAR (project No.\ 277625399, all authors). We acknowledge financial support by the EU (ERC, TopoGrand, 101040409, JSchmitt), by the DLR with funds provided by the BMWi (Grant No. 50WM2240, FV and MW), and by the Cluster of Excellence ML4Q (EXC 2004/1-390534769, MW and JSchmitt).

\textbf{Data availability} \par 

Data presented in this manuscript are available in the Zenodo repository, \url{https://doi.org/10.5281/zenodo.10571407}

\section{Figures}

\begin{figure}
\includegraphics[width=\textwidth]{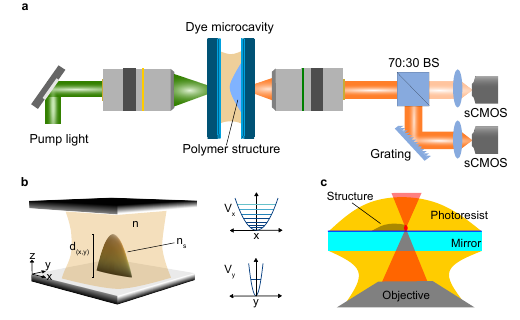}
\caption{\textbf{Dye-filled microcavity experimental setup and cavity mirror nanostructing.}
\textbf{a}, Dye-filled microcavity experimental setup. {The photon gas} is created by pumping the intracavity dye solution using a laser beam spatially shaped using a spatial light modulator (SLM), {and} focused with a 10$\times$ objective into the microcavity. The cavity consists of two {plane} mirrors, with a polymer structure printed on one of them to provide a potential for the photons. The cavity emission is sampled using an imaging objective and subsequently analysed either spatially or spectrally. \textbf{b}, The polymer structure (refractive index $n_{s}$) surrounded by dye solution (refractive index $n$) results in a potential for the trapped photon gas. \textbf{c}, Direct laser writing scheme, using a focused laser beam to polymerize the photoresist on top of the mirror surface.
}
\end{figure}

\begin{figure}
\includegraphics[width=\textwidth]{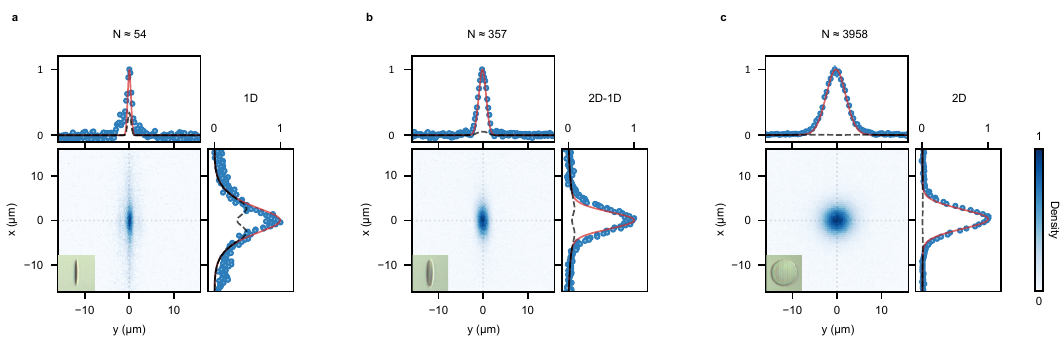}
\caption{
\textbf{Spatial density distribution.}
Density distribution of photons in the quantum degenerate regime in a 1D (\textbf{a}, $\Lambda =22$), 2D-1D (\textbf{b}, $\Lambda =5$) and 2D (\textbf{c}, $\Lambda =1$) harmonic oscillator potentials, respectively. Microscope images of the corresponding polymer structures on the cavity mirror are shown as insets. The dashed lines indicate the position of the cuts through the center of the cloud along the horizontal and vertical axis shown in the side and upper panel. The dashed black line in the cut panels shows the contribution from thermal modes, the solid red line the contribution of the ground mode, {showing the macroscopic contribution from the ground mode in all panels}. For the theoretical expectations we assume a Bose-Einstein distribution of the population within the modes, with the total photon number of $N = 54 $ (1D), $N = 357$ (2D-1D) and $N = 3958$ (2D), respectively. The visible deviation in the 1D case is attributed to the emission of free-space modes which are excited at the rim of the potential.
}

\end{figure}

\begin{figure}
\includegraphics[width=\textwidth]{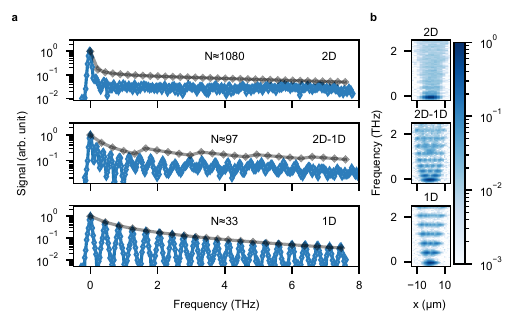}
\caption{
\textbf{Photon gas spectroscopy.}
\textbf{a}, The integrated spectrum of the cavity fluorescence for 2D ($\Lambda=1$), 2D-1D ($\Lambda=5$) and 1D ($\Lambda=22$) harmonic oscillator potential and the black curve is the expected Bose-Einstein distribution of photons {at $T=\SI{300}{\kelvin}$}, evaluated at the measured position of the harmonic oscillator modes. \textbf{b}, Imaging the cavity emission dispersed by a grating onto a camera (raw spectra) allows one to image the shape of the first few modes. For the 2D-1D case the energy states associated with the first excited mode of the tightly confined dimension appear at around {the} 5th mode. The photon number $N$ in each case is chosen in the quantum degenerate regime, such that all modes are visible, i.e. the population $N_0$ in the ground mode does not significantly exceed that of the excited modes, i.e. $N_0/N \approx 0.3$, $0.18$ and $0.3$ for 2D, 2D-1D and 1D respectively. Additional spectra deep in the quantum degenerate regime for 1D are shown in the online Methods.
}
\end{figure}

\begin{figure}
\includegraphics[width=\textwidth]{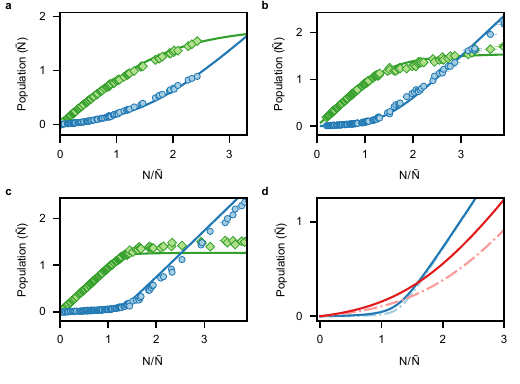}
\caption{
\textbf{Ground vs Excited modes population.}
\textbf{a} - \textbf{c}, show the population in the ground mode (blue dots) and excited modes (green diamonds), for the 1D case ($\Lambda =22$, \textbf{a}, the 1D-2D case ($\Lambda =5$, \textbf{b}) and the isotropic 2D potential in \textbf{c} for a varying total number of photons. The solid lines give the theory expectations assuming a Bose-Einstein distribution within the modes. Panel \textbf{d}, compares the theory expectations for the population in the ground mode for a 1D (red) and a 2D (blue) potential with an equal number of energy levels, with trap depth of $1.2 {k_{B}T}$ (solid lines) for finite size system and quasi-infinite system (dash-dot lines) with a depth of $10 {k_{B}T}$. One clearly observes that the effects from the finite size of the system are weaker than the effects of the dimensional crossover. For better comparison, {the horizontal axis for} each data set is scaled to the photon number $\tilde{N}$, with $\tilde{N}=628$, 64 and 23 photons for the 2D, the 2D-1D and the 1D harmonic oscillator potentials, see text. Error bars showing the statistical standard deviations are of the order of the marker size. {Data are presented as mean values +/- SD.} 
}
\end{figure}

\begin{figure}
\includegraphics[width=\textwidth]{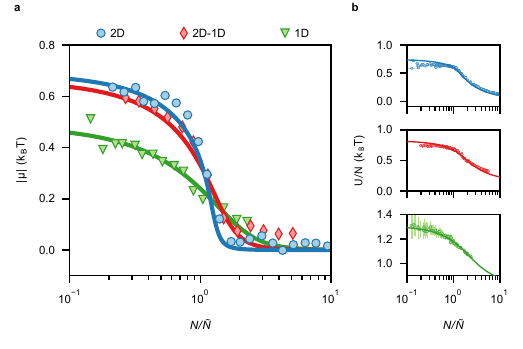}
\caption{
\textbf{Caloric properties.}
The change from a phase transition in 2D to a crossover in 1D is visible in the chemical potential. \textbf{a}, {Symbols} represent the measured change in the absolute of chemical potential $|\mu|$ in the units of thermal energy {($k_\mathrm{B}T$)}, from 1D to 2D (1D, 2D-1D and 2D) harmonic oscillator potentials as function of the normalised photon number $N/\tilde{N}$, where the zero-point energy is set to zero. Solid lines are theory expectations for corresponding harmonic oscillator potentials. \textbf{b}, The measured internal energy per particle (photon), in units of {$k_\mathrm{B}T$}, as a function of the normalised photon number $N/\tilde{N}$, from top to bottom, 2D, 2D-1D and 1D harmonic oscillator potential, with their corresponding theory expectations in solid lines. Error bars show statistical standard deviations and {data are presented as mean values +/- SD. }
}
\end{figure}

%


\end{document}